%% file: main.tex
\documentclass[11pt]{article}

\usepackage[letterpaper,margin=1in]{geometry}
\usepackage[T1]{fontenc}
\usepackage[utf8]{inputenc}
\usepackage{amsmath,amssymb,amsthm,mathtools}
\usepackage{aliascnt}
\usepackage{microtype}
\usepackage{enumitem}
\usepackage{mleftright}
\usepackage{booktabs}
\usepackage{float}
\usepackage{xcolor}
\usepackage[authoryear]{natbib}
\usepackage[colorlinks=true,linkcolor=magenta,citecolor=teal,urlcolor=magenta]{hyperref}
\usepackage{algorithm}
\usepackage{algpseudocode}
\usepackage{tikz}
\usepackage{nicefrac}
\usetikzlibrary{arrows.meta,calc,positioning}

\setlist[itemize]{leftmargin=1.6em,itemsep=0.25em,topsep=0.25em}
\setlist[enumerate]{leftmargin=1.8em,itemsep=0.25em,topsep=0.25em}
\allowdisplaybreaks
\usepackage[capitalize,noabbrev]{cleveref}

\usepackage{xcolor}
\definecolor{myOrange}{RGB}{235, 105, 92}
\definecolor{myMaroon}{RGB}{128, 0, 0}
\definecolor{maroon}{rgb}{0.5,0.0,0.0}
\definecolor{heavymaroon}{rgb}{0.3,0.0,0.0}

\definecolor{mygreen}{RGB}{160, 200, 140}

\usepackage{hyperref}
\hypersetup{
colorlinks=true,
linkcolor=myOrange,
citecolor=myMaroon, filecolor=cyan,
urlcolor=mygreen,
hypertexnames=false
}

\makeatletter
\let\oldalglinenumber\alglinenumber
\renewcommand{\alglinenumber}[1]{%
  \protected@edef\@currentlabel{#1}%
  \oldalglinenumber{#1}%
}
\makeatother

\input{notation}

\title{Settling the Support Complexity of Correlated Equilibria}

\author{
Ashkan Soleymani\footnote{Work done in part during an internship at RBC Borealis.}\\
MIT\\
\texttt{ashkanso@mit.edu}
\and
Patrick Jaillet\\
MIT\\
\texttt{jaillet@mit.edu} \\
\and
Gabriele Farina\\
MIT\\
\texttt{gfarina@mit.edu} 
\and
Akbar Rafiey\\
RBC Borealis, NYU\\
\texttt{akbar.rafiey@rbc.com}
}

\date{}

\begin{document}
\maketitle

\begin{abstract}
We study the support complexity of approximate correlated equilibria (CE) in
two-player games, that is, the minimum number of pure action profiles that must
be assigned positive probability to approximate a correlated equilibrium of a
two-player matrix game. For $\varepsilon$-approximate coarse correlated equilibria,
constant approximation is possible with support $O(\log N/\varepsilon^2)$ in
$N\times N$ games. For correlated equilibria, the best known upper bound was
$O(\log^2 N/\varepsilon^4)$
\citep{babichenko2014simple,babichenko2017empirical},
but whether this bound was tight remained open.
We establish a matching lower bound, resolving the open question of
\citet{ko2017bounds}. Specifically, we show that, already for two-player
zero-sum games, for every $1\leq k\leq N^{1/2}$ the worst-case
correlated-equilibrium gap achievable by a distribution supported on at most
$k$ pure profiles is
\[
  \Theta\mleft(
    \min\left\{1,\frac{\sqrt{\log N}}{k^{1/4}}\right\}
  \mright),
\]
which implies a $\Theta(\log^2 N/\varepsilon^4)$ worst-case support bound.
The proof builds on a probabilistic-method construction based on random
sign matrices.
\end{abstract}

\section{Introduction} \label{sec:intro}

A correlated equilibrium, introduced by
\citet{aumann1974subjectivity,aumann1987correlated}, is a joint distribution
over action profiles from which no player can improve by applying a
recommendation-dependent deviation function.  Exact correlated equilibria exist by
the theorem of \citet{hart1989existence}.  Unlike a Nash equilibrium, which
requires players to randomize independently, a correlated equilibrium may use
arbitrary correlation between the players' actions.  This additional freedom leads
to a major computational difference.  In an explicitly represented normal-form
game, an exact correlated equilibrium can be found in polynomial time by linear
programming, whereas computing a Nash
equilibrium in a two-player bimatrix game is PPAD-complete
\citep{chen2009settling,daskalakis2009complexity}. In this work, we analyze the difference between correlated and Nash equilibria from a different angle: whether correlation also leads to substantially
sparser equilibrium representations.  More precisely, \emph{how many pure action
profiles are necessary and sufficient to support an \(\eps\)-correlated equilibrium
in the worst case?}

This question is part of a long line of work on sparse equilibrium
representations.  In zero-sum games, \citet{althofer1994sparse} and
\citet{lipton1994simple} showed that each player has an $\eps$-optimal
minimax strategy with support $O(\log N/\eps^2)$.  This small-support
approach was extended from zero-sum games to general bimatrix games by
\citet{lipton2003playing}, who showed that every such game admits an
$\eps$-Nash equilibrium in which each player has support
$O(\log N/\eps^2)$.  Approximate Carath\'eodory theorems provide more
general tools for obtaining sparse convex representations, with applications
to sparse approximate equilibria in games
\citep{barman2015approximating,mirrokni2017tight}.

It is important here to distinguish the support of a mixed strategy from the
support of a correlated distribution.  For Nash equilibrium and minimax
strategies, support is measured separately for each player.  If two mixed
strategies \(x\) and \(y\) each have support size at most \(s\), then their
independent product \(x\otimes y\) is a joint distribution supported on at most
\(s^2\) action profiles.  In contrast, a correlated equilibrium is itself a joint
distribution \(\mu\), and its support size is the number of action profiles
\((i,j)\) with \(\mu_{ij}>0\).  Correlation can in principle couple two
large-support marginals using far fewer than the product of their support sizes.
Thus, lower bounds on the support of each player's mixed strategy do not directly
give corresponding quadratic lower bounds on the joint support of a correlated
equilibrium.

For coarse correlated equilibrium, where deviations must be chosen independently
of the recommendation, standard sampling gives joint support
\(O(\log N/\eps^2)\) in every two-player \(N\)-action game.  More generally, if
an equilibrium notion is specified by a finite deviation class \(\mathcal D\),
the standard sampling argument gives support
\(O(\log |\mathcal D|/\eps^2)\).  For CCE in a two-player \(N\)-action game,
there are only \(2N\) relevant constant deviations.  For CE, however, a player
may choose a different deviation for each recommendation.  These deviations are
described by mappings, also called ``swap functions'',
\(f:[N]\to[N]\) and \(g:[N]\to[N]\).  There are \(N^N\) such mappings for each
player, so the direct sampling argument only gives
\(\log |\mathcal D|=\Theta(N\log N)\).  A different argument is therefore needed
to obtain polylogarithmic support. Despite this obstacle, polylogarithmic support is in fact achievable.

For correlated equilibrium, \citet{babichenko2014simple,babichenko2017empirical}
showed that every two-player \(N\)-action game admits an
\(\eps\)-correlated equilibrium with support
\(O(\log^2 N/\eps^4)\).  Before our work, the best known lower bound was
\(\Omega(\log N/\eps^2)\), inherited from support lower bounds for
approximate minimax strategies in zero-sum games
\citep{klein2015number}.  Indeed, the marginals of an
\(\eps\)-correlated equilibrium in a zero-sum game are
\(O(\eps)\)-optimal minimax strategies, and the support of either marginal
cannot exceed the joint support.

This left a quadratic gap between the known upper and lower bounds.  In
particular, it remained possible that correlation could substantially reduce
the joint support needed for approximate equilibrium.  Two approximate
minimax strategies, each with support \(O(\log N/\eps^2)\), give an
\(\eps\)-correlated equilibrium with joint support
\(O(\log^2 N/\eps^4)\) by taking their product. The previously known lower bound with explicit
dependence on $\eps$, however, was $\Omega(\log N/\eps^2)$
joint support.

We close this gap and show that correlation cannot provide such a
reduction of support size in the worst case.  The upper bound of
\citet{babichenko2014simple,babichenko2017empirical} is tight already for
zero-sum games.  For every \(1 \le k \le N^{1/2}\), there exists a zero-sum
\(N\times N\) game in which every distribution supported on at most \(k\)
profiles has correlated-equilibrium gap
\[
\Omega\mleft(
    \min\left\{1,\frac{\sqrt{\log N}}{k^{1/4}}\right\}
  \mright).
\]
This settles the support-size gap highlighted by \citet{ko2017bounds}.  In
particular, the worst-case support complexity of approximate correlated
equilibrium is \(\Theta(\log^2 N/\eps^4)\).  Thus, in zero-sum games, the
product of two low-support approximate minimax strategies already achieves
the optimal worst-case joint support up to constant factors.

More precisely, given a bimatrix game $G=(U,V)$ with
$U,V\in\mathbb{R}^{N\times N}$ and a joint distribution
$\mu\in\Delta([N]^2)$, we define its correlated-equilibrium gap
$\Gap_G(\mu)$ as the largest expected gain that either player can obtain
from a recommendation-dependent deviation,
\[
  \Gap_G(\mu)\coloneqq
  \max\mleft\{
    \max_{f:[N]\to[N]}
    \sum_{i,j}\mu_{ij}\bigl(U_{f(i),j}-U_{ij}\bigr),
    \max_{g:[N]\to[N]}
    \sum_{i,j}\mu_{ij}\bigl(V_{i,g(j)}-V_{ij}\bigr)
  \mright\}.
\]
Thus, $\mu$ is an $\eps$-correlated equilibrium if
$\Gap_G(\mu)\leq\eps$.

Our main result gives a tight lower bound on the CE gap achievable by
distributions with small joint support.

\begin{theorem}\label{thm:main}
Let $\mathcal{Z}_N$ denote the set of bimatrix games with matrices
$U,V\in[0,1]^{N\times N}$.  There exists a universal constant $c>0$ and
$N_0\in\N$ such that, for every $N\geq N_0$ and every integer
$1\leq k\leq N^{1/2}$,
\[
  \sup_{G\in\mathcal{Z}_N}
  \inf_{\substack{\mu\in\Delta([N]^2)\\ |\supp(\mu)|\leq k}}
  \Gap_G(\mu)
  \geq
  c\min\left\{1,\frac{\sqrt{\log N}}{k^{1/4}}\right\}.
\]
\end{theorem}

Combined with the upper bound of
\citet{babichenko2017empirical}, \Cref{thm:main} determines the worst-case
CE gap up to universal constant factors.  In particular, for
$1\leq k\leq N^{1/2}$,
\[
  \sup_{G\in\mathcal{Z}_N}
  \inf_{\substack{\mu\in\Delta([N]^2)\\ |\supp(\mu)|\leq k}}
  \Gap_G(\mu)
  =
  \Theta\mleft(
    \min\left\{1,\frac{\sqrt{\log N}}{k^{1/4}}\right\}
  \mright).
\]
Moreover, our lower bound already holds for the subclass of
\emph{constant-sum} games, which are equivalent to zero-sum games for our
purposes.  We provide a technical overview and proof sketch in
\Cref{sec:lower}, and the full proof in \Cref{sec:detailed_proofs}.

Equivalently, \Cref{thm:main} determines the worst-case support complexity
of approximate correlated equilibrium.

\begin{corollary}\label{cor:support-complexity}
Let $S_N(\eps)$ denote the minimum support size of an
$\eps$-approximate correlated equilibrium in the worst-case $N\times N$
game,
\[
  S_N(\eps)\coloneqq
  \sup_{G\in\mathcal{Z}_N}
  \min\mleft\{
    |\supp(\mu)| :
    \mu\in\Delta([N]^2),\ 
    \Gap_G(\mu)\leq\eps
  \mright\}.
\]
There exist universal constants $c_0,C_0,c_1,C_1>0$ such that, whenever
$N$ is sufficiently large and
\[
  \numberthis{eq:epsilon-range}
  C_0\sqrt{\log N}\,N^{-1/8}\leq\eps\leq c_0,
\]
we have
\[
  c_1\frac{\log^2 N}{\eps^4}
  \leq
  S_N(\eps)
  \leq
  C_1\frac{\log^2 N}{\eps^4}.
\]
In particular, for every sufficiently small fixed constant $\eps>0$,
we have $S_N(\eps)=\Theta_\eps(\log^2 N)$.
\end{corollary}

This shows that correlation does not reduce the worst-case joint support
below the quadratic scale obtained by taking the product of two
low-support approximate minimax strategies.  It also separates CE from CCE
in terms of joint support.  In the regime of
\Cref{cor:support-complexity}, approximate CCE requires only
$\Theta(\log N/\eps^2)$ joint support in the worst case, whereas approximate
CE requires $\Theta(\log^2 N/\eps^4)$. The result also has a direct implication for learning.
For example, this rules out bandit learning algorithms that use a sparse empirical distribution of realized play as a certificate of convergence to an approximate CE.

\section{Preliminaries and Notation}\label{sec:preliminaries}

For a positive integer $N$, let $[N]=\{1,\ldots,N\}$.  We identify a joint distribution on $[N]^2$ with a nonnegative matrix $\mu\in\R_+^{N\times N}$ satisfying $\sum_{i,j}\mu_{ij}=1$.  Its support is
\[
  \supp(\mu)\coloneqq\{(i,j):\mu_{ij}>0\}.
\]
We write $\mu_{i,\ast}$ for the $i$-th row vector and $\mu_{\ast,j}$ for the $j$-th column vector.  Thus, $\mu_{i,\ast}$ records the probability mass on profiles whose row recommendation is $i$, and $\mu_{\ast,j}$ records the analogous mass for column recommendation $j$.  All logarithms are natural and universal constants may change from line to line.

\paragraph{Approximate Correlated Equilibrium}

A two-player normal-form game has row-player and column-player payoff matrices $U,V\in[0,1]^{N\times N}$. A row swap function is a map $f:[N]\to[N]$; after being recommended row $i$, the row player plays $f(i)$.  A column swap function is a map $g:[N]\to[N]$ defined similarly.

\begin{definition}\label{def:gap}
For a game $G=(U,V)$ and a joint distribution $\mu\in\Delta([N]^2)$, define its swap regrets and CE gap as
\begin{align*}
  \Reg_R^G(\mu)
  &\coloneqq \max_{f:[N]\to[N]} \sum_{i,j}\mu_{ij}\bigl(U_{f(i),j}-U_{ij}\bigr),\\
  \Reg_C^G(\mu)
  &\coloneqq \max_{g:[N]\to[N]} \sum_{i,j}\mu_{ij}\bigl(V_{i,g(j)}-V_{ij}\bigr),\\
  \Gap_G(\mu)&\coloneqq\max\{\Reg_R^G(\mu),\Reg_C^G(\mu)\}.
\end{align*}
The distribution $\mu$ is a swap-function $\eps$-correlated equilibrium if $\Gap_G(\mu)\leq\eps$.
\end{definition}

The identity swap function is available, so both regrets are nonnegative.  For $\eps=0$, this is the standard correlated equilibrium condition.  For $\eps>0$, it is the ``global'' swap-function definition used in \citet{babichenko2014simple,babichenko2017empirical}, which is stronger than bounding each individual recommendation-to-action gain separately by $\eps$.

\paragraph{Zero-Sum and Constant-Sum Games}

Let $\mathcal{Z}_N$ be the set of bimatrix games with payoff matrix $U\in[0,1]^{N\times N}$ for the row player and payoff matrix $V\in[0,1]^{N\times N}$ for the column player.  For our lower bounds, we focus on the constant-sum subclass in which $V=\one\one^\top-U$.  Subtracting constants from payoffs does not affect deviations, so this subclass is equivalent, for our purposes, to zero-sum games.

For our lower bounds, we focus on games constructed from a sign matrix $A\in\{-1,+1\}^{N\times N}$ by defining the normalized constant-sum game
\[
\numberthis{eq:sign-normalization}
  U=\frac{\one\one^\top+A}{2},\qquad
  V=\frac{\one\one^\top-A}{2}.
\]
Let $\Reg_R^A$ and $\Reg_C^A$ denote swap regrets in the sign zero-sum game with row payoff $A$ and column payoff $-A$.  This normalization divides regrets by two.
\[
\numberthis{eq:divide-two}
\Reg_R^G(\mu)=\frac12\Reg_R^A(\mu),\qquad
  \Reg_C^G(\mu)=\frac12\Reg_C^A(\mu).
\]

Finally, we use the symbol $G_N(k)$ to denote the worst-case CE approximation of a distribution with support $k$ across the bimatrix games in $\mathcal{Z}_N$. Specifically, for $1\leq k\leq N^2$, we define
\[
  G_N(k)
  \coloneqq \sup_{G\in\mathcal{Z}_N}\ \inf_{\substack{\mu\in\Delta([N]^2)\\ |\supp(\mu)|\leq k}} \Gap_G(\mu).
\]

\section{Proof Sketch and Technical Overview}\label{sec:lower}

In a zero-sum game $(A, -A)$, the $\pm \sum_{ij} A_{ij}\mu_{ij}$ terms in the definition of swap regret cancel out when summing the row and column regret, leading to
\[
\numberthis{eq:sum-cancellation}
  \Reg_R^A(\mu)+\Reg_C^A(\mu)
  =\sum_i \max_r\ip{A_{r,\ast}}{\mu_{i,\ast}}
   +\sum_j \max_c\ip{-A_{\ast,c}}{\mu_{\ast,j}}.
\]
Since $\Reg_R^A(\mu), \Reg_C^A(\mu)$ are nonnegative, showing that right-hand side is large is sufficient to conclude that the CE gap is large.

The cancellation \eqref{eq:sum-cancellation} reduces showing that the CE gap is large to controlling the maxima on the right-hand side. In particular, we will seek a randomized matrix $A$ that ensures that the maxima must be large with high probability whenever $\mu$ is too sparse.  The hard instance $A$ is obtained by drawing a matrix with independent Rademacher entries
and then normalizing it to a constant-sum game with payoffs in \([0,1]\).  

\begin{remark}
A related random-sign construction appears in \citet{azrieli2013lipschitz} to rule out pure approximate Nash equilibria in many-player Lipschitz games. Their argument needs
the random matrix to control the finite set of pure sign profiles, whereas our setting
requires a substantially stronger uniform property.  We must simultaneously control all
sparse nonnegative vectors with arbitrary weights, since the rows and columns of a sparse
correlated distribution can be any such vectors.  This continuous family cannot be handled
by a direct union bound over pure profiles and leads to the sparse norming and interpolation
arguments developed below.
\end{remark}

\subsection{The Effect of Rademacher Sign Matrices on the CE Gap}\label{sec:rade_norming}

For a fixed vector $z$ and an integer parameter
$1\leq p\leq N$, it will be crucial to determine the scale at which a Rademacher sum
$\sum_i \eps_i z_i$ is positive with probability at least
$\exp(-\Theta(p))$. This scale falls somewhat between $\|z\|_1$ and
$\sqrt{p}\,\|z\|_2$ where large coordinates in $z$ must be penalized individually, while
the remaining spread-out mass is penalized via its Euclidean norm.

For \(z\in\R_{+}^N\), let $
  z_1^*\geq z_2^*\geq\cdots\geq z_N^*\geq 0 $
be the nonincreasing rearrangement of \((z_i)_{i=1}^N\).  For an integer
\(1\leq p\leq N\), define the interpolation scale
\[
\numberthis{eq:kappa-def}
  \kappa_p(z)
  \coloneqq
  \sum_{\ell\leq p} z_\ell^*
  +
  \sqrt p
  \left(\sum_{\ell>p}(z_\ell^*)^2\right)^{1/2}.
\]
We also write
\[\numberthis{eq:tau-sigma}
  \tau_p(z)\coloneqq\sum_{\ell\leq p} z_\ell^*,
  \qquad
  \sigma_p(z)\coloneqq
  \left(\sum_{\ell>p}(z_\ell^*)^2\right)^{1/2},
\]
so that \(\kappa_p(z)=\tau_p(z)+\sqrt p\,\sigma_p(z)\).

The first term, $\tau_p(z)$, accounts for the $p$ largest coordinates of
$z$. Their contributions can be aligned by forcing the corresponding
Rademacher signs to be $+1$, an event of probability
$\exp(-\Theta(p))$.  The second term,
\(\sqrt p\,\sigma_p(z)\), accounts for the remaining tail. A Rademacher sum
over the tail has standard deviation \(\sigma_p(z)\), and an upward fluctuation
at \(\sqrt p\) standard deviations again has probability \(\exp(-\Theta(p))\).
Thus \(\kappa_p\) determines the scale at which a sparse nonnegative vector can be
detected by one of exponentially many sign tests. More specifically, we show a one-sided lower bound on the upper-tail probability of an i.i.d. Rademacher
vector at the interpolation scale $\kappa_p$, in the spirit of \citet{montgomery1990distribution}. The following technical lemma is proved in \Cref{app:norming}.
\begin{lemma}\label{lem:rademacher-tail}
There exist universal constants $c_*,C_*>0$ and an integer $p_0$ such that, for every $p\geq p_0$, every $z\in\R_+^N$, and independent Rademacher signs $\varepsilon_1,\ldots,\varepsilon_N$,
\[
  \Prb\left[\sum_{i=1}^N\varepsilon_i z_i\geq c_*\kappa_p(z)\right]
  \geq e^{-C_*p}.
\]
\end{lemma}

Armed with the one-sided Rademacher tail bound, we can now show that a random sign matrix detects every sparse nonnegative vector at the interpolation scale $\kappa_p$.  The theorem below proves precisely this property for random sign matrices. %

\begin{theorem}[Sparse Rademacher norming]\label{thm:sparse-norming}
There exist universal constants $a,c,c'>0$ and $N_0\in\N$ such that for all $N > N_0$ the following holds.  Let
\[
  p\coloneqq\lfloor a\log N\rfloor,
  \qquad
  K\coloneqq\lfloor N^{1/2}\rfloor.
\]
If $A\in\{-1,+1\}^{N\times N}$ has independent uniform entries, then with probability at least $1-2\exp(-c'N^{3/4})$, simultaneously for every nonnegative $z\in\R_+^N$ with $|\supp(z)|\leq K$,
\[
  \max_{r\in[N]}\ip{A_{r,\ast}}{z}&\geq c\kappa_p(z),\\
  \max_{c\in[N]}\ip{-A_{\ast,c}}{z}&\geq c\kappa_p(z).
\]
\end{theorem}

We defer the proof of the theorem to \Cref{app:norming}. The proof uses a covering argument, which is handled by a combination of the Paley-Zygmund inequality and a union bound over the largest coordinates.

\subsection{Mixed-Norm Interpolation Bound} \label{sec:mixed_norm}

The random norming result in \Cref{thm:sparse-norming} shows that the payoff matrix detects each sparse
row or column of a candidate distribution, at least at the scale of $\kappa_p$.  To turn this into a regret
lower bound, we need a deterministic statement saying that a sparse joint
distribution cannot make all of its row and column vectors small at this scale.
This is the point where swap-function correlated equilibrium differs from
one-sided or coarse notions as the same probability matrix is \emph{simultaneously
conditioned by all rows and columns}.

The next lemma ensures that every
$k$-supported probability matrix has nontrivial total $\kappa_p$-mass across
its row and column vectors. We remark the exponent $1/4$ in the result, which then propagates to the main theorem.

\begin{lemma}[Interpolation lower bound]\label{lem:row-column}
Let $\mu\in\R_+^{N\times N}$ satisfy $\norm{\mu}_1=1$ and $|\supp(\mu)|\leq k$.  For any integer $p\geq1$, set
\[
  S_R\coloneqq\sum_i\kappa_p(\mu_{i,\ast}),\qquad
  S_C\coloneqq\sum_j\kappa_p(\mu_{\ast,j}),
  \qquad
  S\coloneqq S_R+S_C.
\]
Then
\[
  S\geq \min\left\{\frac12,\frac{\sqrt p}{k^{1/4}}\right\}.
\]
\end{lemma}

The proof, in \Cref{app:mixed_norm_obstruction}, is fundamentally a pruning argument over rows and columns. Recall that $\kappa_p = \tau_p + \sqrt p\,\sigma_p$, where the $\tau_p$-part measures the largest entries and the $\sigma_p$-part measures the residual $\ell_2$-mass. Suppose the total
$\kappa_p$-mass $S$ is small.  The $\tau_p$-parts then imply that deleting
the $p$ largest entries in every row and in every column removes only small
total mass, so a constant amount of $\ell_1$-mass remains.  On the remaining
matrix, the $\sigma_p$-parts imply that both the sum of row $\ell_2$-norms and
the sum of column $\ell_2$-norms are small.  Blei's mixed $4/3$ inequality
converts these two-sided $\ell_2$ bounds into a small $\ell_{4/3}$-norm~\citep{popa2013blei}.
Finally, $k$-sparsity converts $\ell_{4/3}$ back to $\ell_1$ with loss
$k^{1/4}$. Since the remaining matrix still has constant $\ell_1$-mass, this
is impossible unless $S\gtrsim \sqrt p/k^{1/4}$, up to the trivial constant
lower bound.

\subsection{Connecting the Pieces}

We now show that in a random sign game, every sparse probability matrix has enough total $\kappa_p$-mass so that 
no $k$-supported candidate with gap below the desired bound can exist.

\begin{theorem}\label{thm:random-hard}
There exist universal constants $c,c'>0$ such that the following holds for all sufficiently large $N$.  Let $A\in\{-1,+1\}^{N\times N}$ be a random sign matrix and let $G_A$ be the normalized constant-sum game~\eqref{eq:sign-normalization}.  With probability at least $1-2\exp(-c'N^{3/4})$, every joint distribution $\mu$ with support size $k\leq N^{1/2}$ satisfies
\begin{equation}\label{eq:random-hard}
  \Gap_{G_A}(\mu)
  \geq
  c\min\left\{1,\frac{\sqrt{\log N}}{k^{1/4}}\right\}.
\end{equation}
\end{theorem}

\begin{proof}
Condition on the event in \Cref{thm:sparse-norming}.  Let $p=\lfloor a\log N\rfloor$, and let $\mu$ be supported on at most $k\leq N^{1/2}$ pure profiles.  Every row vector $\mu_{i,\ast}$ and every column vector $\mu_{\ast,j}$ is nonnegative and has support at most $k\leq K$.  Applying \Cref{thm:sparse-norming} to all these vectors and using the cancellation identity~\eqref{eq:sum-cancellation} gives 
\begin{align*}
  \Reg_R^A(\mu)+\Reg_C^A(\mu)
  &\geq c_0\sum_i\kappa_p(\mu_{i,\ast})
      +c_0\sum_j\kappa_p(\mu_{\ast,j}) \\
  &= c_0 S.
\end{align*}
This simultaneous application is justified because, on the event from
\Cref{thm:sparse-norming}, both norming inequalities hold for every
nonnegative vector $z\in\mathbb{R}_+^N$ with
$|\supp(z)|\leq K$. Since each row vector $\mu_{i,*}$ and each column
vector $\mu_{*,j}$ has support at most $k\leq K$, the corresponding
inequalities apply to all of these vectors on the same event.

By \Cref{lem:row-column},
\[
  S\geq \min\left\{\frac12,\frac{\sqrt p}{k^{1/4}}\right\}.
\]
Since both swap regrets are nonnegative,
\[
  \max\{\Reg_R^A(\mu),\Reg_C^A(\mu)\}
  \geq \frac{c_0}{2}\min\left\{\frac12,\frac{\sqrt p}{k^{1/4}}\right\}.
\]
Finally, $p=\Theta(\log N)$ and the normalization~\eqref{eq:divide-two} divides regrets by two.  Absorbing constants $c := c_0\sqrt{a}/4$ proves~\eqref{eq:random-hard}.
\end{proof}

We therefore arrive at the main result of the paper.

\begin{repeatthm}{thm:main}[Extended version]
There exists a universal constant $c>0$ and $N_0\in\N$ such that, for every $N\geq N_0$ and every integer $1\leq k\leq N^{1/2}$,
\[
  G_N(k) \ge c\min\left\{1,\frac{\sqrt{\log N}}{k^{1/4}}\right\}.
\]
Moreover, the lower-bound instance may be
obtained by sampling a random sign matrix \(A\in\{-1,+1\}^{N\times N}\) with
independent entries and applying the normalization~\eqref{eq:sign-normalization}.
With probability at least \(1-2\exp(-c'N^{3/4})\), the resulting game satisfies
the displayed lower bound simultaneously for all \(1\leq k\leq N^{1/2}\).
\end{repeatthm}
\begin{proof}

\Cref{thm:random-hard} shows that, with probability at least
$1-2\exp(-c'N^{3/4})$, a random sign matrix produces a normalized
constant-sum game for which every distribution supported on at most
$k\leq N^{1/2}$ profiles has gap at least the right-hand side
of~\eqref{eq:random-hard}. Since this probability is positive for all
sufficiently large $N$, such a game exists by the probabilistic method.
This proves the lower bound in the definition of $G_N(k)$.
\end{proof}

\section{Detailed Proofs} \label{sec:detailed_proofs}

We begin with the statements and proofs of the auxiliary lemmas used in our derivations.

\begin{lemma}[Paley--Zygmund]\label{lem:paley}
If $Y\geq0$ has finite second moment and $0<\theta<1$, then
\[
  \Prb\mleft[Y\geq \theta \E[Y]\mright]
  \geq (1-\theta)^2\frac{\E [Y]^2}{\E [Y^2]}.
\]
\end{lemma}

\begin{lemma}\label{lem:nets} 
For every $d\geq1$ and $0<\delta\leq1$, the Euclidean unit sphere in $\R^d$, and hence its intersection with the nonnegative orthant, has a $\delta$-net of cardinality at most $(1+2/\delta)^d$.
\end{lemma}

\begin{lemma}\label{lem:log-cosh}
For every real $u$ with $|u|\leq1/2$,
\[
  \frac{u^2}{3}\leq\log\cosh u\leq\frac{u^2}{2}.
\]
\end{lemma}
\begin{proof}
For the upper bound, let $h(u)=u^2/2-\log\cosh u$.  The function is even, $h(0)=h'(0)=0$, and
\[
  h''(u)=1-\operatorname{sech}^2u=\tanh^2 u\geq0.
\]
Thus $h(u)\geq0$.  For the lower bound, let $g(u)=\log\cosh u-u^2/3$.  Again $g$ is even and $g(0)=g'(0)=0$.  On $|u|\leq1/2$,
\[
  g''(u)=\operatorname{sech}^2u-\frac23
  \geq \operatorname{sech}^2(1/2)-\frac23>0.
\]
Hence $g$ is convex on this interval and minimized at zero.
\end{proof}

\subsection{Proofs for \Cref{sec:rade_norming}} \label{app:norming}

In this section we prove \Cref{thm:sparse-norming}.  We start with elementary properties of $\kappa_p$ that become useful in the covering argument.

\begin{lemma}\label{lem:kappa-lipschitz}
For all $z,w\in\R^N$ and all $p\geq1$,
\[
  \kappa_p(z)\geq \norm{z}_2,
  \qquad
  |\kappa_p(z)-\kappa_p(w)|\leq 2\sqrt p\,\norm{z-w}_2.
\]
\end{lemma}

\begin{proof}
Let $z^{\mathrm{top}}$ retain the $p$ largest coordinates of $z$ in magnitude and let $z^{\mathrm{tail}}=z-z^{\mathrm{top}}$.  Then
\[
  \tau_p(z) = \norm{z^{\mathrm{top}}}_1 \geq \norm{z^{\mathrm{top}}}_2,
  \qquad
  \sqrt p\,\sigma_p(z) = \sqrt{p} \, \norm{z^{\mathrm{tail}}}_2 \geq \norm{z^{\mathrm{tail}}}_2.
\]
Thus $\kappa_p(z)\geq \norm{z^{\mathrm{top}}}_2+\norm{z^{\mathrm{tail}}}_2\geq\norm z_2$.

For the Lipschitz bound, observe that
\[
  \tau_p(z)=\max_{|\mathcal{S}|\leq p}\norm{z_{\mathcal{S}}}_1.
\]
Therefore, since $\tau_p$ is subadditive,
\[
  |\tau_p(z)-\tau_p(w)|\leq \tau_p(z-w)\leq \sqrt p\,\norm{z-w}_2.
\]
Also,
\[
  \sigma_p(z)=\inf_{v:|\supp(v)|\leq p}\norm{z-v}_2,
\]
because the best $p$-sparse approximation in Euclidean norm is obtained by retaining the $p$ largest coordinates.  Distance to a fixed set is $1$-Lipschitz, so
\[
  |\sigma_p(z)-\sigma_p(w)|\leq\norm{z-w}_2.
\]
Adding the two estimates proves the claim. Since $\kappa_p = \tau_p + \sqrt{p}\,\sigma_p$, by the triangle inequality and the two estimates above,
$
|\kappa_p(z) - \kappa_p(w)|
\le |\tau_p(z) - \tau_p(w)| + \sqrt{p}\,|\sigma_p(z) - \sigma_p(w)|
\le \sqrt{p}\,\|z - w\|_2 + \sqrt{p}\cdot\|z - w\|_2
= 2\sqrt{p}\,\|z - w\|_2.
$
\end{proof}

We now prove a one-sided lower bound on the upper-tail probability of an i.i.d. Rademacher vector at the interpolation scale $\kappa_p$, in the spirit of \citet{montgomery1990distribution}.

\begin{repeatlemma}{lem:rademacher-tail}
There exist universal constants $c_*,C_*>0$ and an integer $p_0$ such that, for every $p\geq p_0$, every $z\in\R_+^N$, and independent Rademacher signs $\varepsilon_1,\ldots,\varepsilon_N$,
\[
  \Prb\left[\sum_{i=1}^N\varepsilon_i z_i\geq c_*\kappa_p(z)\right]
  \geq e^{-C_*p}.
\]
\end{repeatlemma}

\begin{proof}
By homogeneity, no normalization is needed. Reorder the coordinates so
that $z_1\geq\cdots\geq z_N\geq0$. Put
$$
  A_0\coloneqq\sum_{i\leq p}z_i,
  \qquad
  B_0\coloneqq\left(\sum_{i>p}z_i^2\right)^{1/2}.
$$
Note that $A_0=\tau_p(z)$ and $B_0=\sigma_p(z)$, so
$\kappa_p(z)=A_0+\sqrt{p}B_0$. If fewer than $p$ coordinates are
nonzero, then the event that all nonzero signs are $+1$ has probability
at least $2^{-p}$ and yields the full sum $\kappa_p(z)$. We may therefore
assume that at least $p$ coordinates are nonzero.

\emph{Case 1: $A_0\geq\sqrt{p}\,B_0$.}
Require the first $p$ signs to be $+1$. The tail sum
$T\coloneqq\sum_{i>p}\varepsilon_i z_i$ is symmetric, so
$\Prb[T\geq0]\geq1/2$. The two events are independent and hence have
joint probability at least $2^{-p-1}$. On their intersection,
$$
  \sum_i\varepsilon_i z_i
  \geq A_0
  \geq \frac12\left(A_0+\sqrt{p}\,B_0\right).
$$

\emph{Case 2: $A_0<\sqrt{p}\,B_0$.}
In this case $B_0>0$, and the largest tail coefficient satisfies
$$
  \max_{i>p}z_i
  \leq z_p
  \leq \frac{A_0}{p}
  <\frac{B_0}{\sqrt{p}}.
$$
Set
$$
  \lambda\coloneqq\frac{\sqrt{p}}{4B_0}.
$$
Then $|\lambda z_i|<1/4$ and $|2\lambda z_i|<1/2$ for every $i>p$.
Let $Y=e^{\lambda T}$. By \Cref{lem:log-cosh},
$$
  \E[Y]
  =\prod_{i>p}\cosh(\lambda z_i)
  \geq
  \exp\left(\frac{\lambda^2B_0^2}{3}\right),
$$
and
$$
  \E[Y^2]
  =\prod_{i>p}\cosh(2\lambda z_i)
  \leq
  \exp\left(2\lambda^2B_0^2\right).
$$
Applying \Cref{lem:paley} with $\theta=1/2$ gives
$$
  \Prb\left[Y\geq\frac12\E[Y]\right]
  \geq
  \frac14\frac{(\E[Y])^2}{\E[Y^2]}
  \geq
  \frac14\exp\left(-\frac43\lambda^2B_0^2\right)
  =
  \frac14e^{-p/12}
  \geq e^{-C_1p},
$$
where $C_1\coloneqq\log 4+1/12$ is universal. On this event,
\begin{align*}
  T
  &\geq \frac{\log\E[Y]-\log 2}{\lambda} \\
  &\geq \frac{\lambda B_0^2}{3}-\frac{\log 2}{\lambda} \\
  &=\left(\frac1{12}-\frac{4\log 2}{p}\right)\sqrt{p}\,B_0.
\end{align*}
After increasing $p_0$ so that $p_0\geq96\log 2$, this implies
$$
  T\geq\frac1{24}\sqrt{p}\,B_0.
$$

Now also require $\varepsilon_1=\cdots=\varepsilon_p=+1$. This event is
independent of the preceding event, since $T$ depends only on the signs
$(\varepsilon_i)_{i>p}$. Therefore, the joint event has probability at
least
$$
  2^{-p}e^{-C_1p}=e^{-C_2p},
  \qquad
  C_2\coloneqq C_1+\log 2.
$$
On this event,
$$
  \sum_i\varepsilon_i z_i
  \geq A_0+\frac1{24}\sqrt{p}\,B_0
  \geq \frac1{24}\left(A_0+\sqrt{p}\,B_0\right).
$$

Thus, in both cases,
$$
  \Prb\left[
    \sum_i\varepsilon_i z_i
    \geq c_*\kappa_p(z)
  \right]
  \geq e^{-C_*p}
$$
for universal constants $c_*>0$ and $C_*>0$, for example
$c_*=1/24$ and $C_*=\max\{2\log 2,C_2\}$.
\end{proof}

We are ready to restate and prove \Cref{thm:sparse-norming}.

\begin{repeatthm}{thm:sparse-norming}[Sparse Rademacher norming]
There exist universal constants $a,c,c'>0$ and $N_0\in\N$ such that for all $N > N_0$ the following holds.  Let
\[
  p\coloneqq\lfloor a\log N\rfloor,
  \qquad
  K\coloneqq\lfloor N^{1/2}\rfloor.
\]
If $A\in\{-1,+1\}^{N\times N}$ has independent uniform entries, then with probability at least $1-2\exp(-c'N^{3/4})$, simultaneously for every nonnegative $z\in\R_+^N$ with $|\supp(z)|\leq K$,
\[
  \max_{r\in[N]}\ip{A_{r,\ast}}{z}&\geq c\kappa_p(z),\numberthis{eq:row-norming}\\
  \max_{c\in[N]}\ip{-A_{\ast,c}}{z}&\geq c\kappa_p(z).\numberthis{eq:column-norming}
\]
\end{repeatthm}

\begin{proof}
    We first establish the row statement~\eqref{eq:row-norming}.  Fix a nonnegative unit vector $z$ with support size at most $K$.  By \Cref{lem:rademacher-tail}, each random row $A_{r,\ast}$ independently satisfies
\[
  \ip{A_{r,\ast}}{z}\geq c_*\kappa_p(z)
\]
with probability at least $q \coloneqq e^{-C_*p}$.  Choose the universal constant $a>0$ small enough that $q\geq N^{-1/4}$ for all sufficiently large $N$.  Hence
\begin{equation}\label{eq:fixed-z-failure}
  \Prb\left[\max_r\ip{A_{r,\ast}}{z}<c_*\kappa_p(z)\right]
  \leq(1-q)^N\leq e^{-N^{3/4}}.
\end{equation}

We now make the estimate uniform.  Fix a support $S\subseteq[N]$ with $|S|=d\leq K$.  Let $\mathbb{S}_{S,+}$ be the nonnegative part of the Euclidean unit sphere supported on $S$.  Put
\[
  \delta_d\coloneqq\frac{c_*}{4(\sqrt d+2c_*\sqrt p)}.
\]
By \Cref{lem:nets}, $\mathbb{S}_{S,+}$ has a $\delta_d$-net $\mathcal{N}_S$ with
\[
  |\mathcal{N}_S|\leq\left(1+\frac{2}{\delta_d}\right)^d.
\]
Taking the union over all supports of cardinality at most $K$, the total number of net points is at most
\begin{equation}\label{eq:net-size}
  |\mathcal{N}|
  \leq \sum_{d=1}^K \binom{N}{d}\left(1+\frac{2}{\delta_d}\right)^d
  \leq \exp(CK\log N)
\end{equation}
for a universal $C$ and all sufficiently large $N$.  Indeed, $\binom{N}{d}\leq(eN/d)^d$, while $\delta_d^{-1}=O(\sqrt d+\sqrt p)\leq O(\sqrt N)$.

Applying~\eqref{eq:fixed-z-failure} to every net point and taking a union bound, the probability that some net point fails is at most
\[
  \exp(CK\log N-N^{3/4})\leq\exp(-N^{3/4}/2)
\]
for all sufficiently large $N$, since $K=N^{1/2}+O(1)$.

Assume now that every net point succeeds.  Let $z\in\mathbb{S}_{S,+}$ and choose $z_0\in\mathcal{N}_S$ with $\norm{z-z_0}_2\leq\delta_d$.  The function
\[
  F_A(w)\coloneqq\max_r\ip{A_{r,\ast}}{w}
\]
is $\sqrt d$-Lipschitz on vectors supported on $S$, because every row restricted to $S$ has Euclidean norm $\sqrt d$.  By \Cref{lem:kappa-lipschitz}, $\kappa_p$ is $2\sqrt p$-Lipschitz.  Therefore
\begin{align*}
  F_A(z)
  &\geq F_A(z_0)-\sqrt d\,\delta_d\\
  &\geq c_*\kappa_p(z_0)-\sqrt d\,\delta_d\\
  &\geq c_*\kappa_p(z)-(2c_*\sqrt p+\sqrt d)\delta_d\\
  &\geq c_*\kappa_p(z)-\frac{c_*}{4}.
\end{align*}
Since $\kappa_p(z)\geq\norm z_2=1$, this is at least $(3c_*/4)\kappa_p(z)$.  Homogeneity extends the conclusion to all nonnegative sparse $z$.

The proof of the column statement~\eqref{eq:column-norming} is identical. The columns of $-A$ are independent uniform sign vectors.  A final union bound over the row and column events proves the theorem, after adjusting constants.
\end{proof}

\subsection{Proofs for \Cref{sec:mixed_norm}} \label{app:mixed_norm_obstruction}

In this section, we formally prove \Cref{lem:row-column}. For a matrix $B=\mleft[b_{ij}\mright]_{i,j=1}^N$, write
\[
\norm{B}_{4/3}\coloneqq\left(\sum_{i,j}|b_{ij}|^{4/3}\right)^{3/4}.
\]
We start with a two-dimensional form of Blei's inequality~\citep{popa2013blei}.

\begin{lemma}[Blei's mixed $4/3$ inequality]\label{lem:blei}
For every nonnegative matrix $B=\mleft[b_{ij}\mright]_{i,j=1}^N$,
\[
  \norm{B}_{4/3}
  \leq
  \left(\sum_i\norm{B_{i,\ast}}_2\right)^{1/2}
  \left(\sum_j\norm{B_{\ast,j}}_2\right)^{1/2}.
\]
\end{lemma}

\begin{proof}
Set $r_i\coloneqq\norm{B_{i,\ast}}_2$ and $c_j\coloneqq\norm{B_{\ast,j}}_2$.  Zero rows or columns may be omitted.  For every remaining entry,
\[
  b_{ij}^{4/3}
  =\left(\frac{b_{ij}^2}{r_i}\right)^{1/3}
   \left(\frac{b_{ij}^2}{c_j}\right)^{1/3}
   (r_i c_j)^{1/3}.
\]
H\"older's inequality with exponents $(3,3,3)$ gives
\begin{align*}
  \sum_{i,j}b_{ij}^{4/3}
  &\leq
  \left(\sum_{i,j}\frac{b_{ij}^2}{r_i}\right)^{1/3}
  \left(\sum_{i,j}\frac{b_{ij}^2}{c_j}\right)^{1/3}
  \left(\sum_{i,j}r_i c_j\right)^{1/3}\\
  &=\left(\sum_i r_i\right)^{2/3}
    \left(\sum_j c_j\right)^{2/3}.
\end{align*}
Raising both sides to the power $3/4$ proves the claim.  
\end{proof}

\begin{lemma}\label{lem:support-interpolation}
If a finite vector or matrix $w$ has at most $k$ nonzero entries, then
\[
  \norm{w}_1\leq k^{1/4}\norm{w}_{4/3}.
\]
\end{lemma}

\begin{proof}
Let $S=\supp(w)$. By H\"older's inequality with conjugate exponents $4/3$ and
$4$,
\[
  \norm{w}_1 = \sum_{a\in S}|w_a|
  \leq \left(\sum_{a\in S}|w_a|^{4/3}\right)^{3/4}|S|^{1/4} = \norm{w}_{4/3}|S|^{1/4} \leq k^{1/4}\norm{w}_{4/3}.
\]
\end{proof}

\begin{repeatlemma}{lem:row-column}[Interpolation lower bound]
Let $\mu\in\R_+^{N\times N}$ satisfy $\norm{\mu}_1=1$ and $|\supp(\mu)|\leq k$.  For any integer $p\geq1$, set
\[
S_R\coloneqq\sum_i\kappa_p(\mu_{i,\ast}),\qquad
S_C\coloneqq\sum_j\kappa_p(\mu_{\ast,j}),
  \qquad
  S\coloneqq S_R+S_C.
\]
Then
\[
  S\geq \min\left\{\frac12,\frac{\sqrt p}{k^{1/4}}\right\}.
\]
\end{repeatlemma}
\begin{proof}
    If $S\geq1/2$, the claim is immediate.  Assume $S<1/2$.

For every row $i$, select the positions of its $p$ largest entries, or all its nonzero entries if there are fewer than $p$.  Do the same for every column.  Let $\nu$ be the matrix obtained from $\mu$ by retaining only entries that were selected by neither their row nor their column.

The total mass deleted by row selections is at most $\sum_i\tau_p(\mu_{i,\ast})\leq S_R$.  The additional mass deleted by column selections is at most $\sum_j\tau_p(\mu_{\ast,j})\leq S_C$.  Hence
\[
\numberthis{eq:nu-mass}
  \norm{\nu}_1\geq1-S>\frac12.
\]
In each row, $\nu$ is supported in the tail remaining after the $p$ largest row entries are removed.  Therefore
\[
  \norm{\nu_{i,\ast}}_2\leq \sigma_p(\mu_{i,\ast})\leq \frac{\kappa_p(\mu_{i,\ast})}{\sqrt p},
\]
and summing over rows gives
\[
\sum_i\norm{\nu_{i,\ast}}_2\leq\frac{S_R}{\sqrt p}.
\]
Similarly,
\[
\sum_j\norm{\nu_{\ast,j}}_2\leq\frac{S_C}{\sqrt p}.
\]
By \Cref{lem:blei},
\[\numberthis{eq:nu-43}
  \norm{\nu}_{4/3}
  \leq \frac{\sqrt{S_RS_C}}{\sqrt p}
  \leq \frac{S}{2\sqrt p}.
\]
Finally, $\nu$ has at most $k$ nonzero entries, so \Cref{lem:support-interpolation} yields
\[
\numberthis{eq:nu-sparse}
  \norm{\nu}_1\leq k^{1/4}\norm{\nu}_{4/3}.
\]
Combining~\eqref{eq:nu-mass}, \eqref{eq:nu-43}, and~\eqref{eq:nu-sparse},
\[
  \frac12<\norm{\nu}_1\leq k^{1/4}\frac{S}{2\sqrt p}.
\]
Thus $S>\sqrt p/k^{1/4}$, as required.
\end{proof}

\section*{Acknowledgement}

We thank an anonymous reviewer for observing that
symmetrizing the construction of \citet{althofer1994sparse}
yields an $\Omega(\log^2 N)$ support lower bound for
sufficiently small constant $\eps$.
Our result additionally establishes the tight $\eps^{-4}$
dependence and shows that the lower bound holds with high
probability for random sign games.

A.S. and P.J. were partially supported by the ONR grant N00014-24-1-2470. G.F. was supported in part by the National Science Foundation awards CCF-2443068 and IIS-2552046, the Office of Naval Research grant N000142512296, and an AI2050 Early Career Fellowship.

\bibliographystyle{unsrtnat}
\bibliography{refs}

\end{document}

%% file: notation.tex
\def\[#1\]{\begin{align*}#1\end{align*}}

\NewDocumentCommand{\numberthis}{om}{%
  \IfNoValueTF{#1}{%
    \refstepcounter{equation}\tag{\theequation}%
  }{%
    \tag{#1}%
  }%
  \label{#2}%
}

\newtheorem{theorem}{Theorem}[section]

\newaliascnt{lemma}{theorem}
\newtheorem{lemma}[lemma]{Lemma}
\aliascntresetthe{lemma}

\newaliascnt{proposition}{theorem}

\aliascntresetthe{proposition}

\newaliascnt{corollary}{theorem}
\newtheorem{corollary}[corollary]{Corollary}
\aliascntresetthe{corollary}

\newaliascnt{claim}{theorem}

\aliascntresetthe{claim}

\theoremstyle{definition}
\newaliascnt{definition}{theorem}
\newtheorem{definition}[definition]{Definition}
\aliascntresetthe{definition}

\newaliascnt{question}{theorem}

\aliascntresetthe{question}

\theoremstyle{plain}
\newaliascnt{remark}{theorem}
\newtheorem{remark}[remark]{Remark}
\aliascntresetthe{remark}

\crefname{theorem}{theorem}{theorems}
\Crefname{theorem}{Theorem}{Theorems}
\crefname{lemma}{lemma}{lemmas}
\Crefname{lemma}{Lemma}{Lemmas}
\crefname{proposition}{proposition}{propositions}
\Crefname{proposition}{Proposition}{Propositions}
\crefname{corollary}{corollary}{corollaries}
\Crefname{corollary}{Corollary}{Corollaries}
\crefname{definition}{definition}{definitions}
\Crefname{definition}{Definition}{Definitions}
\crefname{section}{section}{sections}
\Crefname{section}{Section}{Sections}

\newcommand{\R}{\mathbb{R}}
\newcommand{\N}{\mathbb{N}}
\newcommand{\eps}{\varepsilon}

\newcommand{\Reg}{\operatorname{Reg}}
\newcommand{\Gap}{\operatorname{Gap}}
\newcommand{\supp}{\operatorname{supp}}

\newcommand{\E}{\mathbb{E}}
\newcommand{\Prb}{\mathbb{P}}
\newcommand{\one}{\mathbf{1}}
\newcommand{\ip}[2]{\left\langle #1,#2\right\rangle}
\newcommand{\norm}[1]{\left\lVert #1\right\rVert}

\numberwithin{equation}{section}

\theoremstyle{plain}

\newcommand{\repeatthmname}{}
\newtheorem*{repeatthminner}{\repeatthmname}
\newenvironment{repeatthm}[1]{%
  \renewcommand{\repeatthmname}{Theorem~\ref{#1}}%
  \begin{repeatthminner}%
}{%
  \end{repeatthminner}%
}

\newcommand{\repeatlemmaname}{}
\newtheorem*{repeatlemmainner}{\repeatlemmaname}
\newenvironment{repeatlemma}[1]{%
  \renewcommand{\repeatlemmaname}{Lemma~\ref{#1}}%
  \begin{repeatlemmainner}%
}{%
  \end{repeatlemmainner}%
}

\newcommand{\repeatpropname}{}
\newtheorem*{repeatpropinner}{\repeatpropname}

\newcommand{\repeatcorollaryname}{}
\newtheorem*{repeatcorollaryinner}{\repeatcorollaryname}